\documentclass[a4paper,11pt]{article}

\usepackage{jcappub} 
\def\be{\begin{equation}}
\def\ee{\end{equation}}
\def\ba{\begin{eqnarray}}
\def\ea{\end{eqnarray}}
\def\bce{\begin{center}}
\def\ece{\end{center}}

\usepackage{graphics,color}
\usepackage[T1]{fontenc} 

\title{\boldmath  Spectrum of Supernova Neutrinos
in Ultra-pure Scintillators}

\author[a,b]{C. Lujan-Peschard,}
\author[a]{G. Pagliaroli,}
\author[a,c]{F. Vissani.}

\affiliation[a]{Laboratori Nazionali del Gran Sasso, INFN, Assergi (AQ), Italy}
\affiliation[b]{Departamento de Fisica, DCeI, Universidad de Guanajuato,
  Le\'on, Guanajuato, M\'exico}
\affiliation[c]{Gran Sasso Science Institute, INFN, L'Aquila (AQ), Italy}

\emailAdd{carolup@fisica.ugto.mx}
\emailAdd{giulia.pagliaroli@lngs.infn.it}
\emailAdd{francesco.vissani@lngs.infn.it}

\abstract{
There is a great interest in measuring the non-electronic component of neutrinos from core collapse supernovae by observing, for the first time, also neutral-current reactions.
In order to assess the physics potential of the ultra-pure  scintillators in this respect, we study  the entire expected  energy spectrum in the Borexino, KamLAND and SNO+ detectors.
We examine the various sources of uncertainties in the expectations, and in particular, those due to specific detector features and to the relevant cross sections. 
We discuss the possibility to identify the different neutrino flavors, and we quantify the effect of confusion, due to other components of the energy spectrum, overlapped with the neutral-current reactions of interest.}

\begin{document}
\maketitle
\flushbottom

\section{Introduction}
\label{sec:intro}

A Core Collapse Supernova (SN) releases 99\% of its total energy by
emitting neutrinos of the six flavors. 
The capability to observe the electronic antineutrino 
component of this emission has already been proven by the 
detection of SN1987A neutrinos \cite{KII,IMB,Baksan}.
The very large statistics that we will collect from
the next galactic supernova  will allow us to study the time dependence of the spectrum, specific 
features of the $\bar{\nu}_e$ luminosity and of its average energy \cite{SK, IceCube}. 
The detection of the other neutrinos flavors, however, requires specific detectors and interactions, typically with smaller cross sections. This is true, in particular, for the non electronic component of the spectrum, 
that can be observed only through Neutral Current (NC) interactions. 


During the last years a new generation of ultra-pure liquid
scintillators, Borexino (BRX) \cite{2009BorexinoColl} and
KamLAND (KAM) \cite{2003KamlandColl},  have been operated, obtaining
excellent results thanks to the unprecedented low background
levels reached and the new sensitivity in the very low energy range,
below 1 MeV. They have a particularly good physics potential for the detection supernova 
NC channels, and quite remarkably, the Elastic Scattering (ES) of (anti)neutrinos 
on protons \cite{beato}. It has been argued that the high statistics from this reaction should suffice to constrain the spectra of  the non electronic component for a SN emission, already with the existing detectors  \cite{dasg}. In view of the importance of this conclusion, we would like to reconsider it in this work.

The outline of this paper is as follows. First of all, 
we summarize the available information regarding SN neutrino detection 
in the existing ultra-pure scintillators, and calculate for each of them the total number of expected events as well as their spectral features. We consider the contributions of
all neutrino interaction channels and obtain in this way the   
spectrum of events for a galactic supernova. In this way, we are in the position to evaluate which are 
the capabilities of the present generation of  ultra-pure scintillators 
 to identify and measure the different neutrino flavors.


\section{Emission from a Standard Core Collapse Supernova}

The aim of this work is  to discuss an important question: what we can really see with the existing ultra-pure scintillators and to which extent we can distinguish the different neutrino flavors. With this purpose in mind, we will use very conservative assumptions on the emission model.
We suppose that  the energy radiated in
neutrinos is $\mathcal{E}=3\times 10^{53}$ erg, which is a typical
theoretical value that does not contradict what is found in the most 
complete analyses of SN1987A events \cite{ll,paglia}. 
We also assume that the energy is partitioned in equal amount among the six
types of neutrinos, that should be true within a factor of 2
\cite{Keil:2002in}. 

In agreement with the recent studies, e.g., 
\cite{Tamborra:2012ac}, we consider quasi-thermal neutrinos, 
 each species being characterized by an average energy $\langle E_i\rangle$
 and including a mild deviation from a thermal distribution described by the parameter
$\alpha=3$ for all flavors.
Thus, the neutrino fluence differential in the neutrino energy $E$ is
$$
\Phi_i=\frac{\mathcal{E}_i}{4\pi D^2}\times \frac{E^\alpha e^{-E/T_i}}{ T_i^{\alpha+2} \Gamma(\alpha+2)} \ \ \
  i=\nu_e,\nu_\mu,\nu_\tau,\bar{\nu}_e,\bar{\nu}_\mu,\bar{\nu}_\tau
$$
where the energy radiated in each specie is 
$\mathcal{E}_i=\mathcal{E} f_i$,  with 
$f_i=1/6$ in the case of equipartition, 
and the `temperature' is 
$T_i=\langle E_i\rangle/(\alpha+1)$. In particular, the neutrino and antineutrino 
fluences relevant to NC interactions 
$$
\Phi^{\mbox{\tiny SN}}_\nu=\Phi_{\nu_e}+2 \Phi_{\nu_\mu} \mbox{ and } 
\Phi^{\mbox{\tiny SN}}_{\bar\nu}=\Phi_{\bar{\nu}_e}+2 \Phi_{\bar{\nu}_\mu}
$$
since we suppose that the distribution of the 4 non-electronic species is identical.

The average energies are fixed by the following considerations:
consistent with the simulations in \cite{Tamborra:2012ac} and with the 
findings from SN1987A \cite{ll,paglia}, we set the electron
antineutrino average energy to $\langle E_{\bar{\nu}_e}\rangle=$12 MeV. 
For the average energy of the non-electronic species,
that cannot be seriously probed with SN1987A \cite{paglia}, 
we suppose that the non-electronic temperature is 
30\% higher than the one of $\bar{\nu}_e$: $\langle
E_x \rangle=15.6$ MeV, this is in the upper range of values, 
but still compatible with what is found in \cite{Keil:2002in}. For a
comparison we will consider also the worst case in which the energies
of the non electronic component is equal to the one of the
$\bar{\nu}_e$,  namely  $\langle
E_x \rangle=12$ MeV as showed in very recent simulation \cite{Mueller:2014rna}.
We calculate the electron neutrino average energy by the condition that the
proton (or electron) fraction of the iron core in the neutron star forming is 0.4: 
this gives $\langle E_{\nu_e}\rangle=9.5$ MeV.



Note that the NC reactions are independent from
neutrino oscillations, while for the CC interactions also considering
only the standard oscillation scenario, 
the choice of the mass hierarchy has an important impact on the expectations.
Thanks to the fact that $\theta_{13}$ is large 
(say, larger than about 1 degree) this means that, in normal mass
hierarchy, the survival probability of electron neutrinos and antineutrinos are 
$|U_{e3}^2|$ and $|U_{e1}^2|$ respectively, 
whereas for inverted mass hierarchy, the two values become
$|U_{e2}^2|$ and $|U_{e3}^2|$. 
Thus, the approximate numerical values that we will assume 
in the calculations are
\begin{center}
\begin{tabular}{c|c|c}
   & $P_{\bar\nu_e\to \bar\nu_e}$ & $P_{\nu_e\to \nu_e}$ \\ \hline
   Normal  & 0.7 & 0.0 \\
   Inverted & 0.0 & 0.3 
   \end{tabular}
   \end{center}
The value of $P_{\bar\nu_e\to \bar\nu_e}$ in the case of inverted
hierachy means that what we measure as electronic antineutrinos in
terrestrial detectors, are non-electronic antineutrinos at the
emission in fact; thus, it has a particularly important impact 
on the interpretation of the data. We note also that these numerical values 
would imply that there are only little chances to probe the emission of
electron neutrinos, which are, from the astrophysical point of view,  
the most important type of neutrinos emitted by a supernova.

\begin{figure*}[t]
\bce
\includegraphics[width=0.48\textwidth,angle=0]
{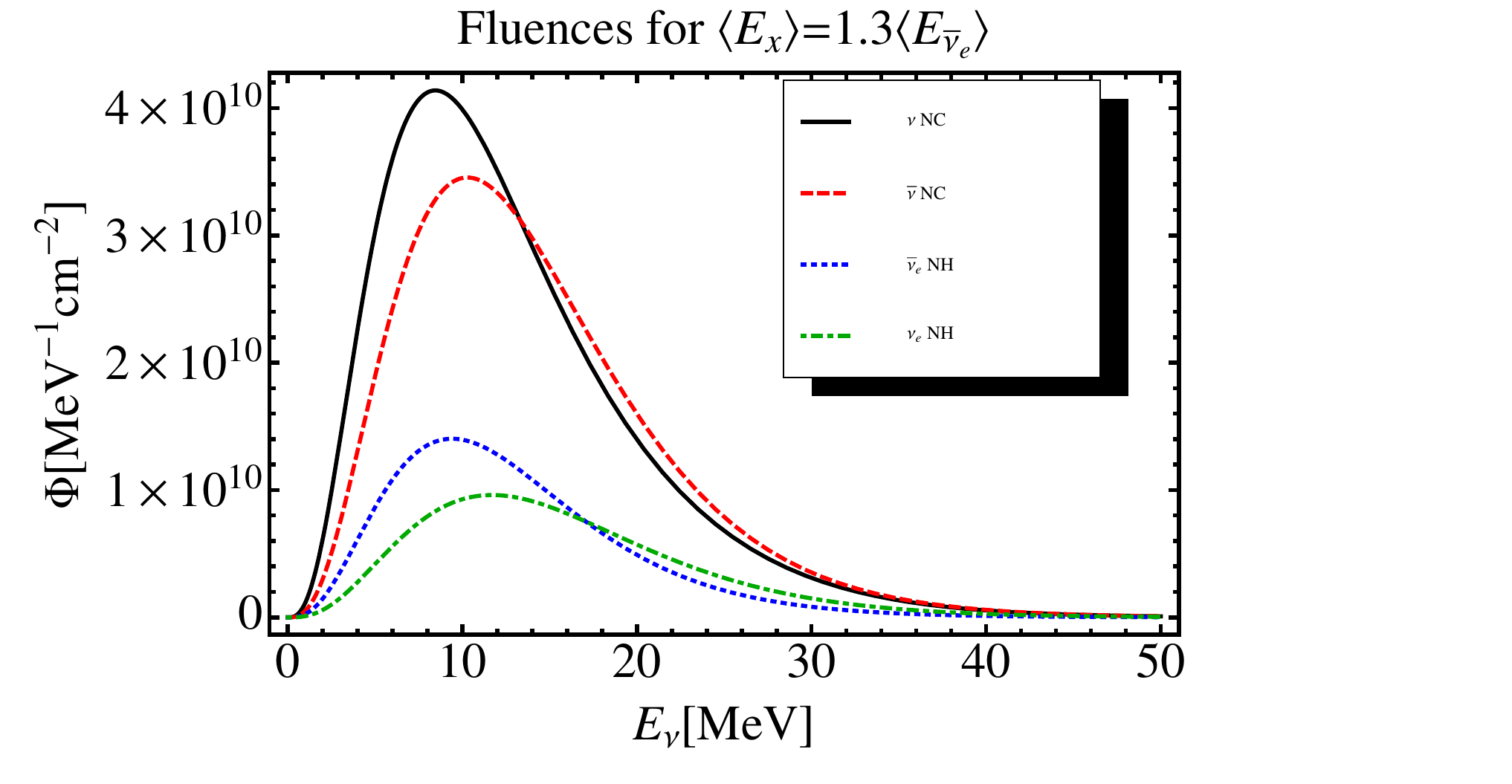}
\includegraphics[width=0.48\textwidth,angle=0]{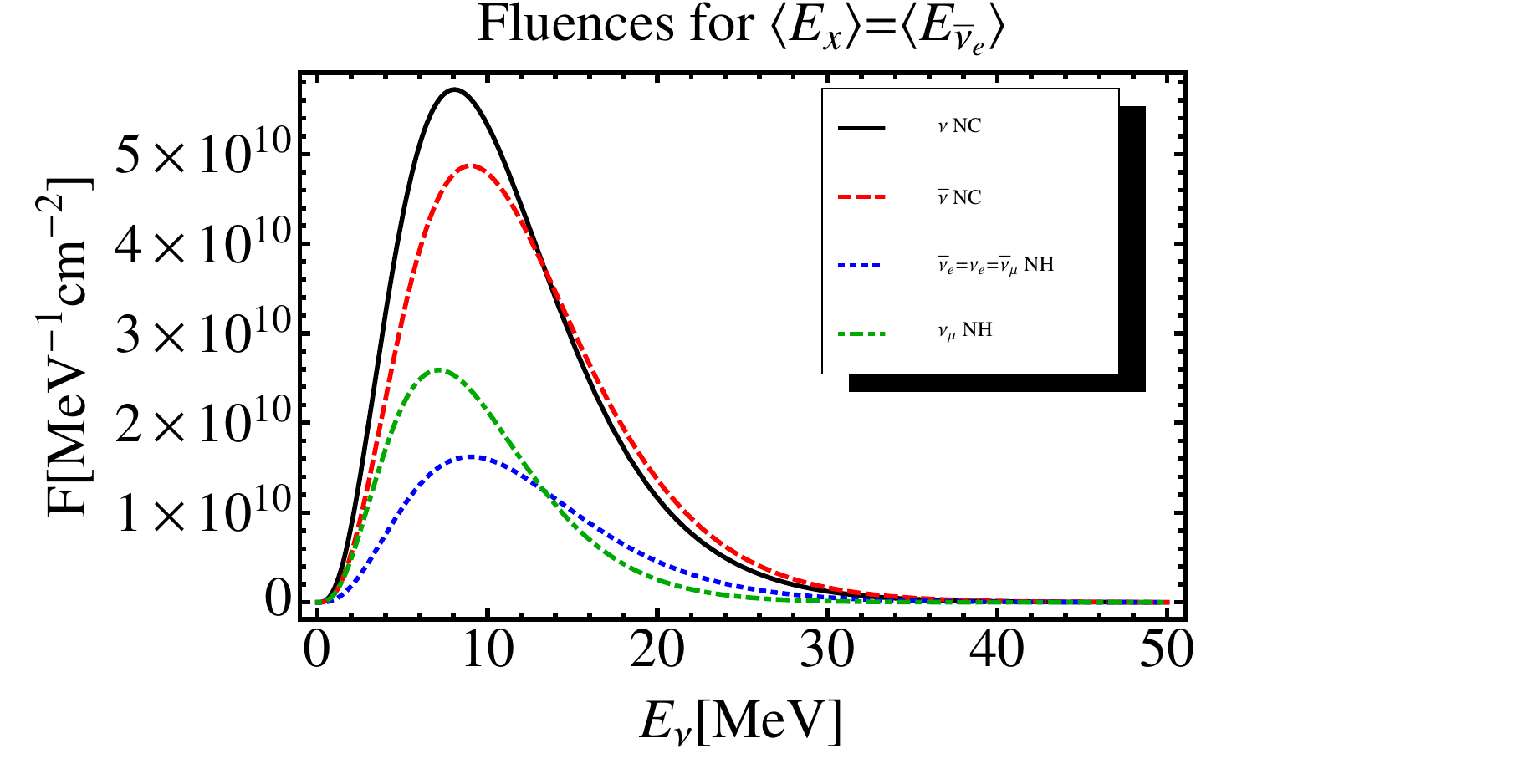}
\ece
\caption{\em Fluences expected for the different neutrinos flavors. 
The blue dotted line shows $\bar{\nu}_e$, the green
dot-dashed line $\nu_e$. We assume standard neutrino
oscillations and Normal mass Hierarchy (NH) for the fluences of each
flavor. The black thick lines and the red dashed ones
show the fluences relevant for the Neutral Current (NC) detection channels of
neutrinos and antineutrinos, respectively.\label{fig1}}
\end{figure*}

In the following we will consider only the case of 
the normal mass hierarchy for definiteness and adding a bit of 
theoretical bias; 
recall however that this hypothesis is immaterial for the discussion of the neutral current events. 
 The total fluences expected to reach the Earth 
under these assumptions are shown in Fig.\ref{fig1} 
for a Supernova exploding at $10$ kpc from us.

\section{Interaction Channels}


In the scintillators and at SN energies, we need consider the several interaction processes:

\paragraph{\bf CC processes.}
Those  involving electronic
antineutrinos are
\begin{itemize} 
\item Inverse Beta Decay (IBD), i.e. $\bar{\nu}_e+p\to n+e^+$;
\item $\bar{\nu}_e + ^{12}C \to ^{12}B + e^+$;
\end{itemize}
while those involving electronic neutrinos are
\begin{itemize} 
\item Proton Knockout $^{12}$C$(\nu, p e^-) ^{11}$C;
\item $\nu_e + ^{12}C \to ^{12}N + e^-$.
\end{itemize}
\paragraph{NC processes.} 
We considered the following channels,
\begin{itemize} 
\item ES on protons, $\stackrel{(-)}\nu p\to \stackrel{(-)}\nu p$;
\item  The 15.11 MeV de-excitation line 
of the  $^{12}C$ nucleus, $\nu ^{12}C \to  \nu ^{12}C^*$;  
\item 
The  Proton Knockout $\stackrel{(-)}\nu+ ^{12}C\to \stackrel{(-)}\nu+ ^{11}B$.
\end{itemize}
Moreover, we consider the ES on electrons, that receives a contribution from both CC and NC.
Let us discuss these reactions in detail.

\subsection*{Detailed description of the cross sections}

The {\bf IBD}, i.e., $\bar{\nu}_e+p\to n+e^+$, 
represents the main signal not only in water 
Cherenkov and also in scintillator detectors. It produces a continuous spectrum due 
to the positrons energy release. The approximated 
kinematic of this reaction connects 
the neutrinos energy with the detected energy through 
$E_\nu=E_d+Q-m_e$ where $Q\simeq1.3$ MeV is the $Q$ 
value of the reaction and $m_e$ is the electron mass.
For the calculations, the IBD cross section reported in
\cite{strumiavissa} was used. 
The {\bf delayed neutron capture} on a proton is characterized by a 
monochromatic $\gamma_{2.2 \text{ MeV}}$ emission. 
The coincidence in a typical time window of about 250 $\mu$s between 
the latter and the prompt signal from the $e^+$ gives a clear 
signature of an IBD event. This means that, in the time integrated events
spectrum, there will be a very high peak around $2.2$ MeV that integrates
the same number of events expected for IBD, reduced by the efficiency of the
tag. The  spectral shape of this peak is due to both the energy resolution of the 
detector and the quenching of the gamma ray energy in the scintillator. In this work, due to lack of information, we neglect the last effect and consider the optimistic case 
in which the width of this peak is only due to the energy resolution.

The {\bf superallowed CC reactions}  $\nu_e+^{12}$C$\to e^-
+^{12}$N and $\bar{\nu}_e+^{12}$C$\to e^+
+^{12}$B present physical thresholds of 
$E_{\nu_{e}}>\!17.3\,$ MeV  and $E_{\bar{\nu}_{e}}\!>\!14.4\,$ MeV respectively. They are detectable
through the prompt leptons $e^{-}$ ($e^{+}$), which give a continuous spectrum. 
Moreover the nucleus of both reactions in the final state, $^{12}$N
and $^{12}$B, are unstable. The former will decay $\beta^+$ to 
$^{12}$C with a half life of $\sim 11$ ms. The latter will decay $\beta^-$ to $^{12}$C with a half life of 
$\sim 20$ ms \cite{N12}. The high energy positrons 
and electrons emitted in these beta decays  
can be observed, giving the possibility to tag these
events. The cross sections used for the evaluation are those 
reported in \cite{Fukugita}.

In the NC channels all neutrino flavors are involved potentially
increasing the number of signal events detected. 

For the {\bf ES on protons} channel the cross section in
\cite{ahrens,xxsec} was used, with a proton strangeness of
$\eta=0.12$. However it is important to stress that the
uncertainty on the number of events expected for this channel 
is not negligible due to the proton structure and amounts to about $20\%$ \cite{cyclotron}. 
To understand the spectral shape of this class of events it is necessary 
to model the quenching factor for protons in  
the scintillators; this accounts for the proton light output and depends on the liquid scintillator 
composition. A detailed description of this factor is given in the next section. 

The cross section for the {\bf superallowed NC reaction} $\nu +^{12}$C $\to \nu+
^{12}$C$^*$ followed by the emission of a monochromatic $\gamma$ at
$15.11$ MeV 
is reasonably well known. It was measured in KARMEN \cite{Karmen}, confirming the
correctness of the calculations as reported in \cite{Fukugita} within an accuracy of 20\%. Future
measurements, most remarkably in OscSNS \cite{oscsns}, claim the
possibility of measuring more than 1,000 events in one year with a
systematic estimated at 5\% level or better. 
The prominent spectral feature of this channel can permit the identification 
of these events, as a sharp peak around $15$
MeV, standing out from the main signal due to IBD. 

The total cross section for {\bf NC proton knockout }
$\nu+{}^{12}\mbox{C}\to \nu + \mbox{p}+^{11}$B  has 
been calculated in \cite{yoshida}, as a part of a network of reactions needed to 
describe the nucleosynthesis of light elements. 
However, the calculation of \cite{haxton} finds a cross
section about 30\% larger, which suggests an error of at least this
order. The neutrino energy has to exceed a pretty high threshold, i.e.  
$E\!>\!\left[(M_B+m_p)^2-M_C^2\right]/(2 M_C)\!\simeq\!15.9$ MeV 
(where we use obvious symbols for the masses of the carbon nucleus, 
of the boron nucleus and of the proton). 
The initial neutrino energy (minus the activation energy, quantified
by the threshold) is shared by the neutrino and the proton in the
final state, $E+M_C\approx E'+T_p+M_B+m_p$
so that the maximum kinetic proton energy $T_p^{\mbox{\tiny max}}$ 
is obtained when the final state neutrino is almost at rest, $E'\approx
0$. The expression for the maximum of the proton kinetic energy is
\begin{equation}
T_p^{\mbox{\tiny max}}=\left[(M^*-m_p)^2-M_B^2\right]/(2M^*),
\end{equation} 
with $M^*=\sqrt{M_C^2+2M_C E}$. 
In view of the smallness of this sample of events, we adopted a very simple procedure to
describe the distribution in the kinetic energy of the proton $T_p$, namely, we
resorted to the pure phase space, that gives ${d\sigma}/{dT_p}\propto
{d\Phi}/{dT_p}\propto\sqrt{T_p}(M^*-m_p-m_B-T_p)^2$. 
We checked that the integral in the proton kinetic
energy of this expression agrees at the level 
of few percent with the theoretical  behaviour of 
the total cross sections as reported in \cite{yoshida}.\footnote{However, a 
word of caution is in order; while the above considerations on phase space
are suggestive, they are just a reasonable way to 
explore of the consequences of this reaction in scintillator detectors: in fact, the distribution in $T_p$ 
of \cite{yoshida} is not available. Certainly, it would be better 
to have a true calculation of the distribution in $T_p$ of this reaction 
(or possibly its parameterization) along with an assessment of the 
theoretical error in the relevant energy range.}

In the {\bf CC knockout proton} reaction on $^{12}$C the outgoing kinetic 
energy is shared between the electron and the proton.  
The maximum kinetic proton energy $T_p^{\mbox{\tiny max}}$ is obtained when the 
final electron is at rest. Similarly to the previous case, 
this is given by
\begin{equation}
T_p^{\mbox{\tiny
    max}}=\left[(M^*-m_p-m_e)^2-M_{C11}^2\right]/(2M^*-2m_e),
\end{equation}
where again  $M^*=\sqrt{M_C^2+2M_C E}$. 
In this case the phase space is
${d\sigma}/{dT_p} \propto {d\Phi}/{dT_p}
\propto\sqrt{T_p}(M^*-m_p-m_{C11}-T_p)^2$. The theoretical cross sections 
reported in \cite{yoshida} and the value estimated from pure 
phase space agree at the level of $\sim 20$\%. 

In the {\bf Elastic Scattering on electrons} all the flavors participate,
but the cross section is slightly different for the different
flavors. The current best measurement of this interaction arises in a 
sample of 191 events \cite{ESe, ESe2}, and quotes 17\% of total error. The error that we estimate in the standard model is instead absolutely negligible for our purposes. 

\subsection*{Numerical formulae}

Let us conclude this section by giving two numerical 
formulas to evaluate easily the main neutral current 
cross sections:\\
An easy-to-implement effective formula for the 
$\nu +^{12}C \to \nu+^{12}C^*$ cross section,
that agrees with \cite{Fukugita} results at better than 1\% in the region below 100 MeV, is
\be
\frac{1}{2}\left(\sigma_\nu+\sigma_{\bar{\nu}}\right)=\frac{G_F^2}{\pi} (E-15.11 \mbox{ MeV})^2 \cdot 
10^{p(E)},
\ee
with $p(E)=\sum_{n=0}^3 c_n \left(E/100\mbox{ MeV}\right)^n$
where $E$ is the incoming neutrino energy, and 
the numerical coefficients are $c_0=-0.146$,   $c_1=-0.184$,  
$c_2=-0.884$, $c_3=+0.233$.  \\
A simple parametrization of the cross section for the ES scattering,
$\nu p \to \nu p$,  assuming that the proton strangeness is 
$\eta=0.12$, is simply 
\be
\frac{1}{2}\left(\sigma_\nu+\sigma_{\bar{\nu}}\right)=G_F^2 E^2 \cdot
10^{q(E)},
\ee
 where $q(E)=-0.333-0.16 (E/100\mbox{ MeV})$.

\section{Description of the Ultra-pure Scintillating Detectors}

We consider the ultrapure liquid scintillators detectors that are running or under construction, 
namely the following three:
Borexino (BRX) \cite{2009BorexinoColl} (0.3 kt of $C_9 H_{12}$) 
in Gran Sasso National Laboratory, Italy, KamLAND 
(KAM) in Kamioka Observatory, Japan \cite{2003KamlandColl} (1 kt of
mixture of $C_{12}H_{20}$(80\%) and $C_9H_{12}$(20\%)) and SNO+ (0.8 kt
of $C_6H_5C_{12}H_{25}$) currently under construction in the SNOLAB facility, located
approximately 2 km underground in Sudbury, Ontario, Canada \cite{SNO}.

We assume that the energy resolution for each detector is a Gaussian with an error described by 
$\sigma(E_d)=A\times \sqrt{E_d/\mbox{MeV}}$ and a different
value of the constant $A$ for each detector. 
The trigger threshold in Borexino is as low as about 200 keV, reaching
full efficiency at $E_d=250$ keV \cite{BorexPRL}. The overall light
collection in Borexino is $\simeq$ 500 photoelectrons (p.e.)/MeV of
deposited energy. The resolution is 
$\simeq$ 5\% at 1 MeV (namely $A=50$ keV).
The trigger efficiency in KamLAND currently reaches 100\% at 350 keV. 
The energy resolution of the KamLAND detector 
can be expressed in terms of the deposited energy as 
$\sim\,6.9\%/\sqrt{E_d(\text{MeV})}$ (i.e., $A=69$ keV) \cite{dasg}.
The energy threshold expected for SNO+ is the optimistic one of $200$
keV and the energy resolution is supposed to be the same as in Borexino.

\begin{table}[t]
  \centering
  \begin{tabular}{|c|c|c|c|}
\hline
 & $a_1$ & $a_2$ & $a_3[\mbox{\small MeV$^{-1}$}]$\\
\hline
BRX  & 0.624 & $-0.175$ & $-0.154$\\
KAM & 0.581 & $-0.0335$ & $-0.207$\\
SNO+ & 0.629  & $ -0.286$ & $-0.163$\\
\hline
\end{tabular}
\label{const}
\caption{Constants appearing in the parametrized formula of the 
quenching function here adopted.}
\end{table}

As we mentioned earlier when the detected particle is a proton, the visible energy  
is only a fraction of the kinetic energy $T_p$,  as described by the `quenching function'. 
Each detector has its own quenching function, that depends on its 
chemical composition; for Borexino detector we consider the quenching
function discussed in \cite{cyclotron}, for KamLAND the one recently
discussed in \cite{KamQuen} and finally for SNO+ the response to
proton in LAB scintillator as measured in \cite{vonKrosigk:2013sa}.  

\begin{figure}[t]
\centering{
\includegraphics[width=0.7\textwidth,
angle=0]{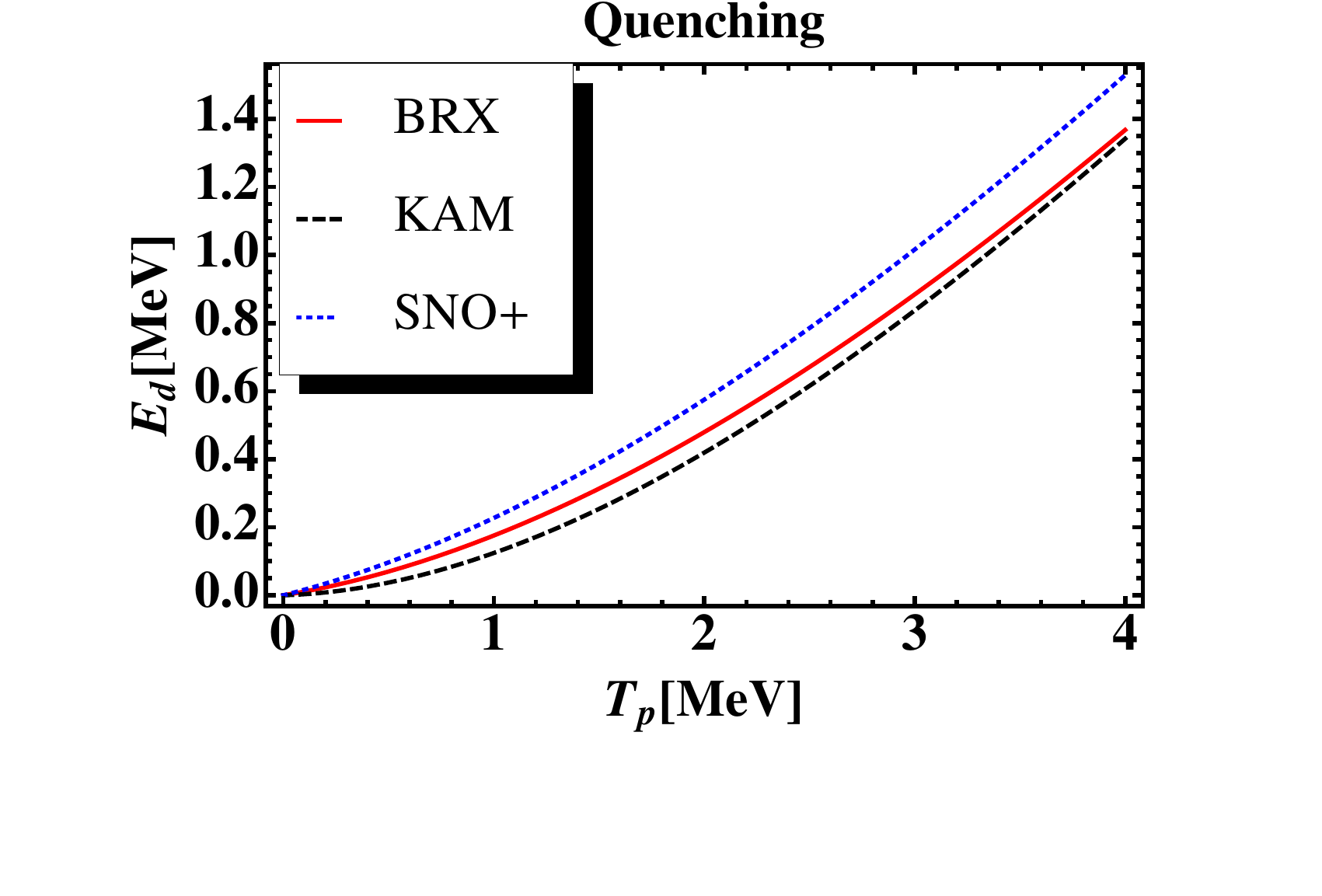}}
\caption{\em Quenching functions used to convert the proton kinetic
  energy in the detectable energy $E_d$ for Borexino (red line),
  KamLAND (black dashed line) and SNO+ (blue dotted line), respectively. \label{fig3}}
\end{figure}

Following \cite{Madey} a simple 
parametrization of the quenching function is 
\begin{equation}
E_d=a_1[1-\exp(a_2+a_3\cdot T_p)]\cdot T_p
\label{quench}
\end{equation}
The values for the constants $a_1$, $a_2$ and $a_3$ that should be used for the 
different detectors are reported in Tab.~\ref{const} and the resulting
functions are shown in Fig.\ref{fig3}.


At this point, we obtain the following important conclusion:
\begin{quote}
in ultrapure scintillators, 
the observation of protons from the NC elastic scattering reaction
allows us to observe only the high energy part of the neutrino spectra.
\end{quote}
In fact, due to the thresholds and to the quenching, 
the protons below a minimum kinetic energy cannot be
detected; this is $0.9$ MeV for SNO+, $1.8$ MeV for Kamland and $1.3$
MeV for Borexino. Thus,  taking into account the kinematical relation between the
proton kinetic energy and the one of incoming neutrinos, we find that  
the elastic scattering on protons is sensitive to neutrino
energies above a threshold of $22$ MeV in the best situation of SNO+, it becomes 
25 MeV for Borexino, and raises to $30$ MeV in the case of Kamland.  In view of these considerations,
one concludes that the exploration of the low energy of the spectrum via neutral currents 
is not possible with the existing ultrapure scientillators.


\section{Results}
\begin{figure*}[t]
\includegraphics[width=0.45\textwidth,angle=0]{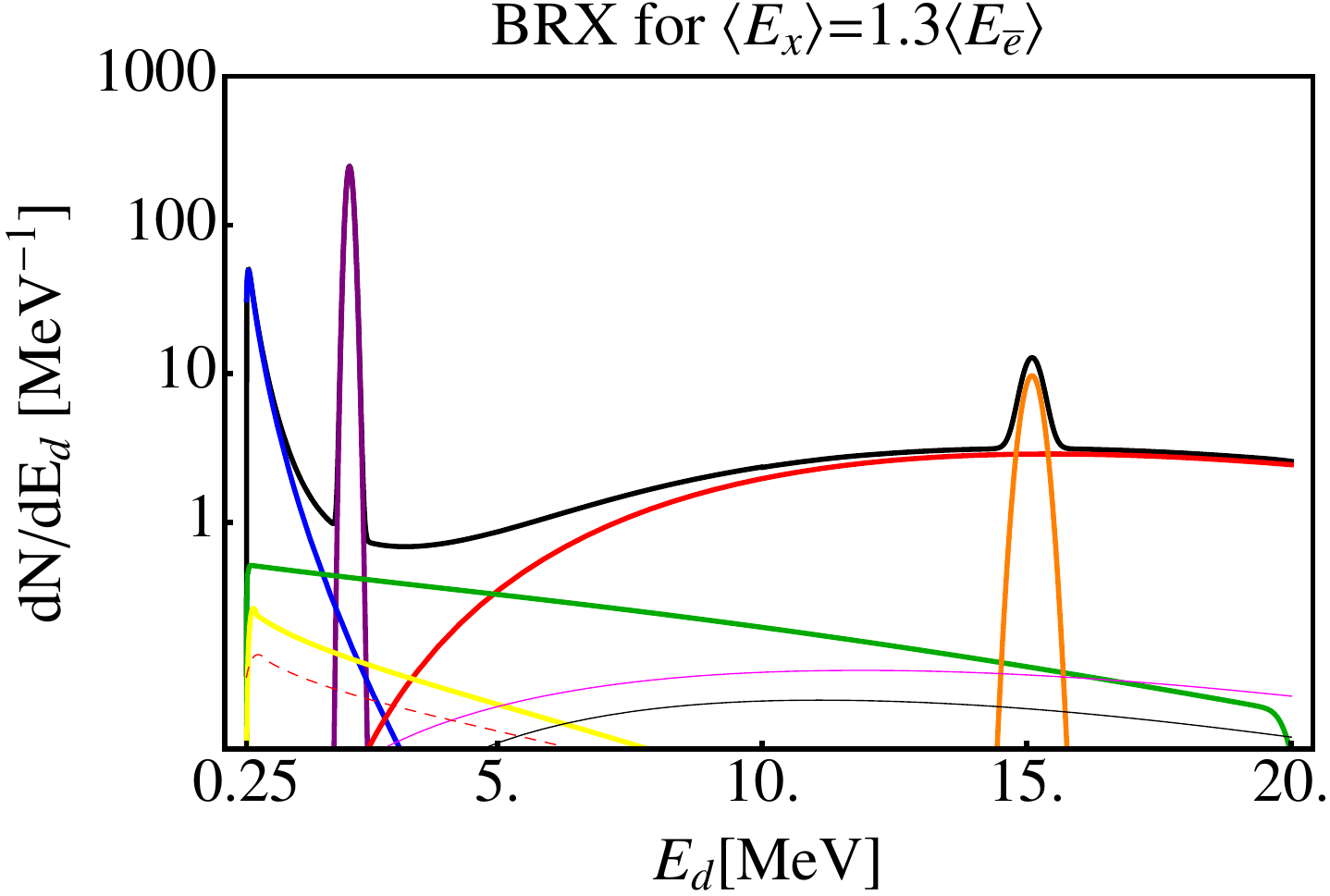} \includegraphics[width=0.45\textwidth,angle=0]{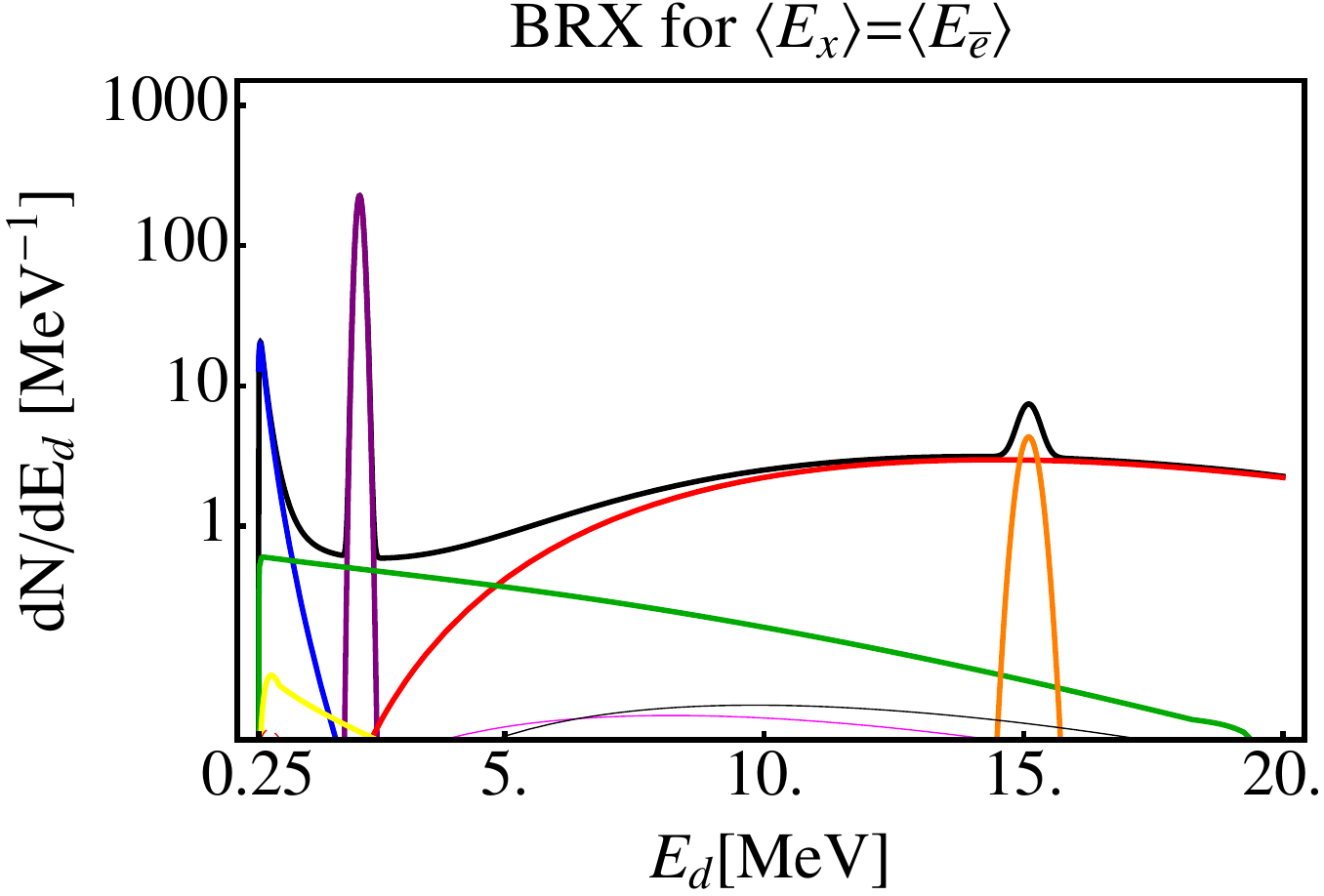}\\
\includegraphics[width=0.45\textwidth,angle=0]{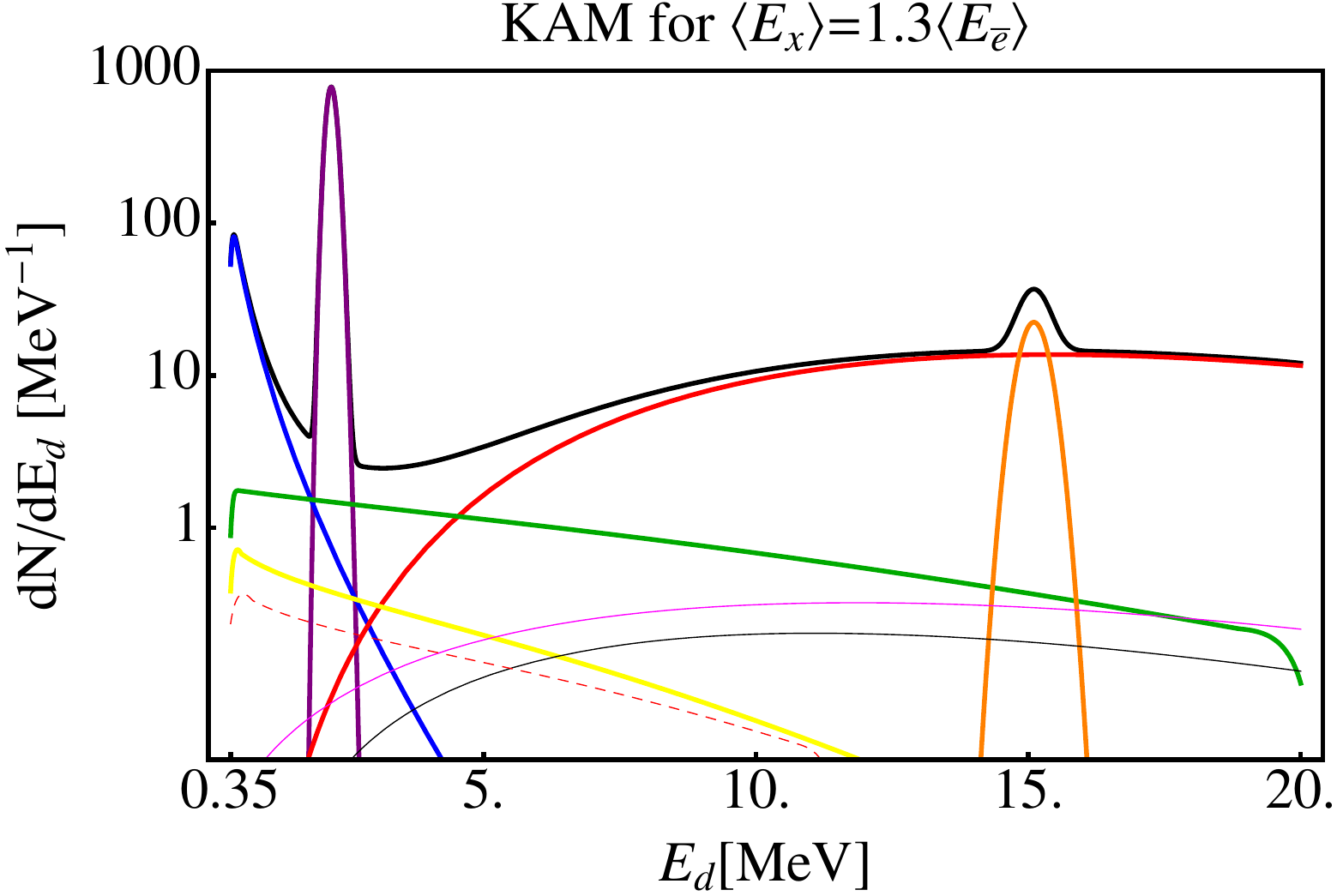} \includegraphics[width=0.45\textwidth,angle=0]{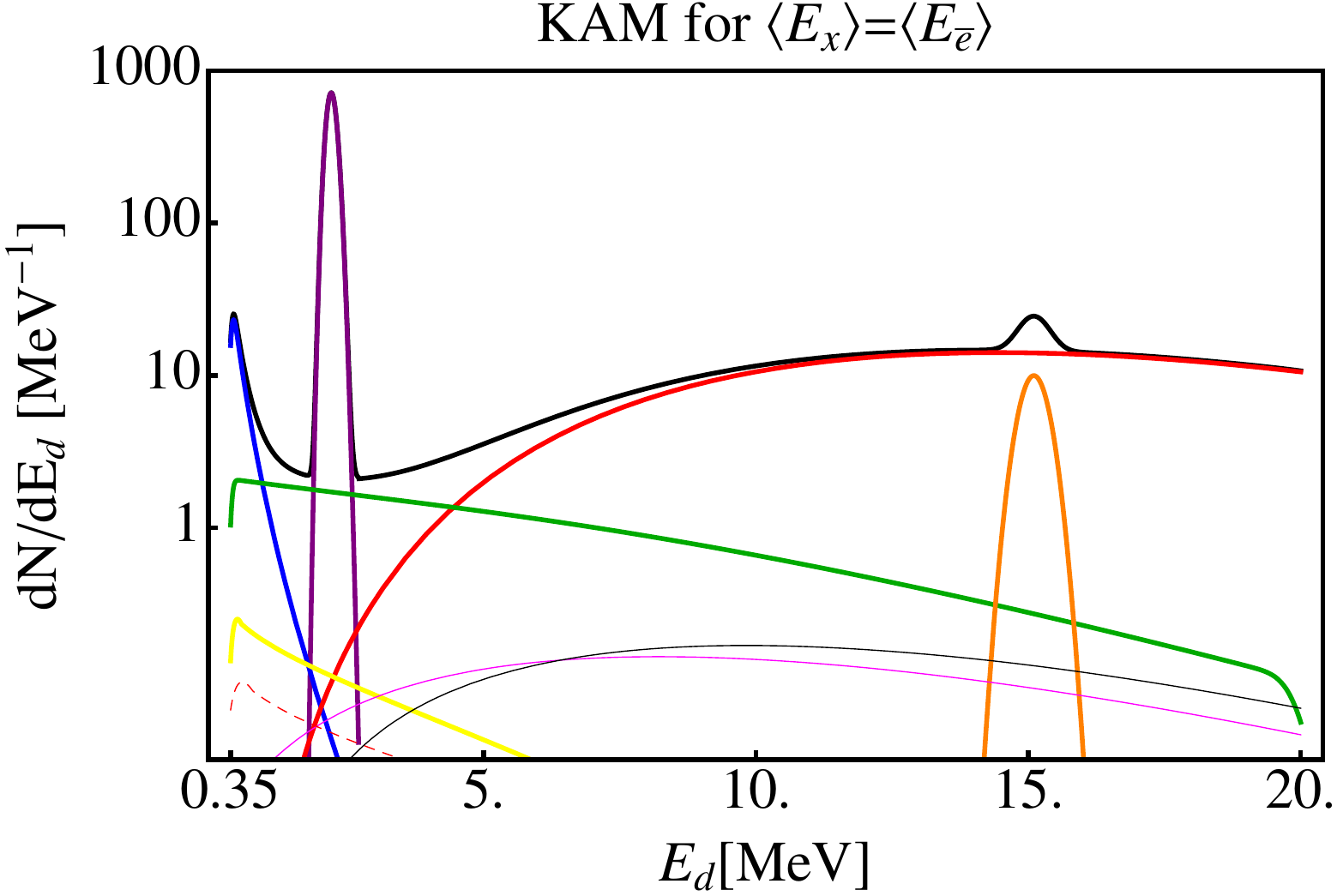}\\
\includegraphics[width=0.45\textwidth,angle=0]{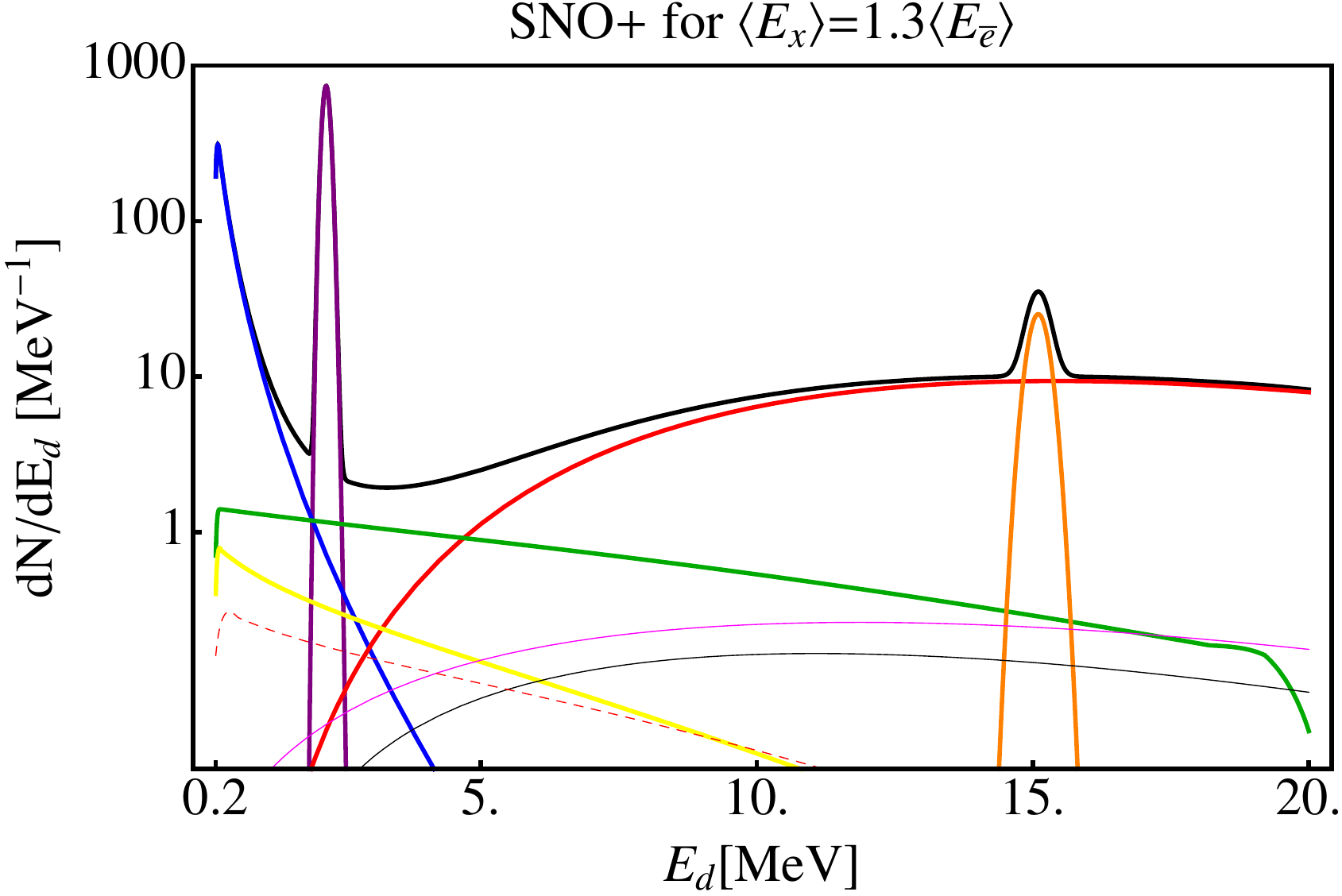} \includegraphics[width=0.45\textwidth,angle=0]{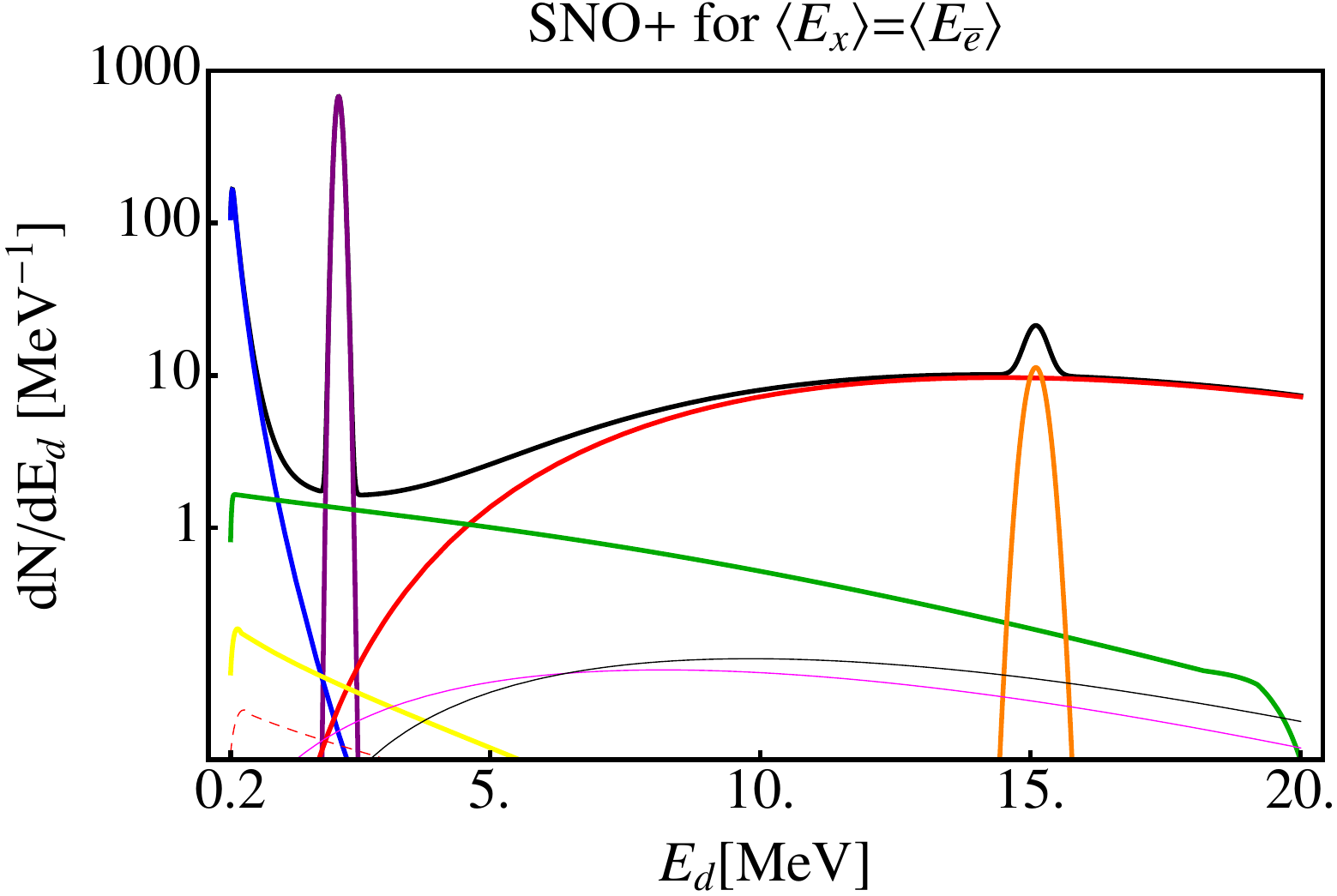}
\caption{\em Theoretical event spectrum in the detectable energy $E_d$ 
in ultrapure scintillators for a Supernova exploding at 10 kpc. The top
panels refer to Borexino detector, the bottom panels to
SNO+, those in the middle to KamLAND. On the left panels, the expectations for the
emission model where energy of non-electronic component is 30\%
higher then the one of $\bar{\nu}_e$; on the right panels, the corresponding
expectations for the case when the non-electronic
components and the $\bar{\nu}_e$ have the same average energy.
See Table~\ref{tab} and the text for the explanation of the individual lines. \label{fig2}}
\end{figure*}

For each detection channel, we estimate the number of expected events and report them in Table~\ref{tab}.  Moreover, we plot the energy distributions of the events, considering the specific features of the ultrapure scintillating detectors in Figure~\ref{fig2}.

The IBD channel (red line) starts to dominate the global signal at $5$ MeV and reaches the maximum around $14$ MeV. The total number
of interactions expected for a supernova located at $10$ kpc is of 
about $54$ events for Borexino, $257$ for Kamland and $176$ for SNO+. 
These results are reported in the first row of Table \ref{tab}. 
The subsequent  gamma from neutron capture gives the peak at 2.2 MeV, 
shown by a purple line. The efficiency of the neutron tag is 
(85$\pm$1)\% in Borexino (see \cite{2010Borexino}), (78$\pm$2)\% in 
KamLAND (see \cite{2003KamlandColl}) and also in SNO+.
The condition for a successful IBD tag \cite{2010Borexino} is that no more
than one interaction occurs during the time between the
IBD interaction and the neutron capture inside a specific volume, namely
\begin{equation} 
R_{IBD}\times \Delta t \Delta V \rho \leq 1
\end{equation}
where $R_{IBD}$ is the rate of IBD events per second and per unit mass, 
$\Delta t$ is
the temporal window of the tag, that we assume to be $\Delta t=2\tau = 512 \mu s$, 
$\Delta V$ is the volume of a sphere with 1 meter of radius,
$\rho$ is the density of the scintillator. 
This is related to the detector mass and to the distance of the supernova by
\begin{equation}
R_{IBD}=\frac{256.5}{T}\cdot\left(\frac{10 \text{kpc}}{D}\right)^2\cdot
\left(\frac{M}{1 \text{kton}} \right),
\end{equation}
where $M$ is the mass of the detector, $D$ is the distance of the SN
and $T$ is the duration of the emission. For example to allow the IBD tag in a detector with the
density of Borexino and 1 kton of mass, considering that $50$\% of the
total emission is expected during the first second \cite{paglia}, 
then the minimum distance of a SN is $D\geq 0.16$ kpc, that is not a severe 
limitation.\footnote{If instead, 
a similar detector but with a $50$ kton mass is considered the
distance becomes $D\geq 1.13$ kpc, that includes several known 
potential Core-Collapse SNe as Betelgeuse and VY Canis Majoris \cite{bet}.}

\begin{table*}
  \centering
  \begin{tabular}{|c|c|c||c|c|c|}
\hline
    Channel & Color code & Signal & BRX & KAM & SNO+\\
    \hline
     $\bar{\nu}_e+p\to n+e^+$ & red  & $e^+$ & 54.1 (49.6) & 256.5  (235.3) & 175.8 (161.2) \\
    $n+p\to D +\gamma_{2.2 \text{ MeV}}$ & purple  & $\gamma$ & 46.0 (42.1) &
  200.1(183.5) &137.1 (125.8)\\
      $\nu +p\to \nu+p$ & blue  & $p$ & 12.7 (3.8) & 29.0 (6.2) & 74.9 (29.2)\\
     $\nu +^{12}C \to \nu+^{12}C^*$ & orange & $\gamma$ & 4.7 (2.1) & 15.0
   (6.7) & 12.3 (5.5)\\
      $\nu +e^-\to \nu+e^-$& green  & $e^-$ & 4.4  (4.5) & 14.8 (15.5) &12.0 (12.4) \\
   $\nu_e+^{12}C\to e^- +^{12}N$ & magenta & $e^-$ & 2.0 (0.7) & 6.4
   (2.1) & 5.3 (1.7)\\
   $\bar{\nu}_e+^{12}C\to e^+ +^{12}B$& black thin & $e^+$ & 1.2 (0.8) &
   3.7 (2.6) & 3.0 (2.1) \\
   $\nu +^{12}C \rightarrow \nu + p + ^{11}B$ & yellow &$p$ & 0.7
   (0.2)& 2.4 (0.6) & 2.1 (0.6)\\
  $\nu_e +^{12}C \rightarrow e^- + p + ^{11}C$ & red dashed &$p$ & 0.5
  (0.1) &1.5 (0.3) & 1.3 (0.2) \\
  
\hline
  \end{tabular}
  \caption{Summary table for all the number of events from the various interaction channels. 
  The result are given for the emission model where the
    energy of non-electronic components is 30\% bigger then the one of
  $\bar{\nu}_e$; in brackets, we indicate the corrispondent values assuming that the
  energies of $\bar{\nu}_e$  and $\nu_x$ are the same. 
The color code (2nd colum) refers to the lines of Figure~\ref{fig2}.}
  \label{tab}
\end{table*}

Let us discuss now the NC elastic proton scattering. This channel dominates the low energy
part of the spectrum, represented with the blue line in Figure \ref{fig2}, 
even if as discussed previously,
this reaction probes only the high energy part of the supernova neutrinos.  
It is evident from Table~\ref{tab} and Fig.~\ref{fig2} that 
the event spectrum depends strongly on the quenching of the proton 
signal. The detector threshold used in the
case of KamLAND is an optimistic value, since we assumed a threshold of $350$ keV,
lower than the one that can be obtained with the current radioactivity
level, i.e., $600$ keV \cite{Tolich}. If this higher threshold is
assumed the ES with protons on KamLAND gives only $17$ signal events. 

It is important to remark that the number of events due to this
channel is also very sensitive to the SN emission parameters;
in fact, as discussed in the end of the previous section, this interaction is sensitive only to the high energy
tail of the SN neutrinos spectra. In
particular, the average energy of the different neutrino flavors have 
gradually been changing in recent years, moving toward lower mean 
values \cite{Janka:2012wk} and toward minor differences between the average 
energies of the different components \cite{Ott:2012mr,Tamborra:2012ac}. 
For comparison we have considered the new paradigm of emission, where 
the average energy of non-electronic flavors is the same as of the
electronic antineutrinos, namely $\langle E_x \rangle=12$ MeV 
\cite{Janka:2012wk,Tamborra:2012ac} and have investigated the two different 
cases to outline the impact on this and the other NC process. 
As shown by the values in brackets in Table \ref{tab}, the expectations in this case are quite meager.

The NC neutral current reaction $\nu +^{12}C \to \nu+
^{12}C^*$ followed by the emission of a monochromatic $\gamma$ is
shown in orange in Figure \ref{fig2}. 
This channel does not require the low energy threshold and  
the efficiency for its detection is 
taken 100\% for all the detectors considered.
However, the possibility of a successful identification is affected by the quality of the  
energy resolution of the detector and by the effectiveness to tag the IBD signal.
These events can be observed if the IBD events are identified through
the correlated neutron capture signal, since they are expected to occur in the same energy region. 
With the assumed energy resolutions we have
that this neutral current reaction can be observed in the energy range
(14-16) MeV. For Borexino,  the number of events due to the IBD signal in the
same range is $5.7$; thus, more than the 50\% of the 
total signal collected in this energy window is due to the IBD
channel, while for KamLAND and SNO+ the IBD signal is 26.9 and 18.4, 
representing about 60\% of the total one. In the case of Borexino, 
if the tagging efficiency is of 85\% as assumed in the plot, 
we expect only 1 event due to IBD not identified, so the uncertainty on the $\gamma_{15.11 \text{MeV}}$ 
signal is reduced to 14\%.

The ESe involves all the flavors of neutrinos and we expect to collect
about $4$ events for Borexino, $15 $ for KamLAND and $12 $ for SNO+. 
Their spectrum is reported with a dark green line in the spectra of Figure \ref{fig2}, 
and 
dominates in the energy region between the ES on protons and the IBD
signals. As we mentioned the cross section for the different flavors
are slightly different; the $\nu_e$ contribution produces half of the events.

The rest of the detection channels have a low signal. All of them show a continuos spectrum, being from  $e^{-}$, $e^{+}$ or protons. 
The two CC superallowed reactions are indicated by a magenta line for the $^{12}N$ final state nucleus
and by a black thin line for the $^{12}B$ one. The decay products of the unstable nucleus are not considered in this plot. For both knockout channels, besides the high energy thresholds, the quenching has to be considered, so the total number of events collected for them is pretty small. The one due to NC is shown in yellow, while the one due to CC is shown with the dashed red line.

\section{Conclusions}

In this paper, we have obtained and discussed the spectrum of supernova neutrino 
events in ultrapure scintillators for a supernova exploding 
at 10 kpc from the Earth. We have examined the capability to distinguish the various   
detection channels and we have quantified the uncertainties in this 
type of detectors. 

As discussed in the introduction, a major reason of specific interest  for a future supernova   is the possibility to observe neutral current interactions of neutrinos. 
We have investigated the three possible reactions of detection 
in ultrapure scintillators, namely: 
1)~the elastic scattering with protons, 
2)~the 15.11 MeV $\gamma$ de-excitation line, 
3)~the proton knockout channel. 
Our conclusions are as follows:

The first reaction is 
characterized by the larger number of expected events in all 
the detectors; however  the
number of detectable events is strongly limited by the energy thresholds. 
The uncertainty on the total number of elastic scattering on protons, due 
to the proton structure, amounts to the 20\%; moreover 
in the same energy region where this reaction can be observed 
there are also the indistinguishable events due to the elastic
scattering with electrons 
and those due to NC and CC proton knockout. In other words, all we can observe is the total 
number of events collected in the energy region from the detector threshold 
to the threshold of the IBD signal, about $1.8$ MeV. 
In this detection window, 
we have found that a fraction of 8\% (Borexino), the 7\% (KamLAND), the 4\%
(SNO+) of the signal is due to the other channels and this uncertainty 
is small but irreducible. We have also seen that, in the case that  the 
the energy of non-electronic neutrinos is low, the 
number of events due to this reaction is too small 
to permit the investigation of the $\nu_x$ spectrum 
at the level discussed in~\cite{dasg}.

The gamma line due to neutrino-induced 
$^{12}$C de-excitation is in principle easier, giving a signal 
at a high energies; its detection does not require 
the extreme performances at very low energies are not needed. 
However,  in the same region of the spectrum where this line is visible 
we will have also  positrons due to IBD reaction; 
thus, the efficiency to tag the concomitant neutron 
will be of crucial importance to   identify cleanly a sample of this NC reaction.
While this concern is not a severe issue for the type of detectors we have considered in this work,
it is much more relevant for future scintillators with a much larger mass and limited performances at low energies.

Finally, we have shown that the proton knockout will give a comparably 
small number of NC events. 


\end{document}